\documentclass[A4,12pt]{article}
\usepackage[cp1251]{inputenc}
\usepackage[T2A]{fontenc}
\usepackage{bm}
\usepackage{amsmath}
\usepackage{amsfonts}
\usepackage{amsmath}
\usepackage{amssymb}
\author{{S.L. Yakovlev \footnote{St Petersburg State University, 199034, St Petersburg, Russian Federation}\footnote{ E-mail:  {s.yakovlev@spbu.ru}} \footnote{The work is supported by RFBR grant 18-02-00492а}}
}

\title{\bf Weak asymptotics of wave function for  $N$-particle system and asymptotic filtering}

\begin{document}

\date{}
\maketitle

\abstract{
\noindent Asymptotic representations for large values of the hyperradius are constructed for the scattering wave function of a system of $ N $ particles considered as a generalized function of angular variable coordinates. The coefficients of the asymptotic representations are expressed in terms of the $N$-particle scattering matrix. The phenomenon of asymptotic filtration is discovered, which consists in the fact that only scattering processes contribute to the leading terms of such an asymptotic representation, in which all particles are free both before and after interaction. The obtained representations are used to construct the correct asymptotics of the partial components of the wave function of $N$ particles in the hyperspherical representation.
}
\\ \\

\section{Introduction}
\noindent
Studying the asymptotic behavior of the wave function for $N$ particles in scattering states at large distances between particles is one of the most difficult problems in quantum scattering theory. To date, this has been most fully implemented for the three-particle system \cite {Merkuriev1971, Yakovlev2016}. In the case of four particles, the asymptotic behavior of the wave function is obtained for configurations in which the distances between all particles are large 
\cite {Yakovlev1990}. A characteristic feature of the representations obtained in these works is the appearance of so-called intermediate terms with respect to a plane wave corresponding to the free motion of particles and a spherical wave corresponding to the resulting expansion of all particles after interaction. These intermediate terms have different orders of asymptotic behavior in different regions of the asymptotic part of the configuration space, corresponding to the processes of successive rescattering of particles. The change in the asymptotic regime in the representations of these terms is described by special functions like the Fresnel integrals \cite {Merkuriev1971, Yakovlev1990}. For systems with a larger $ (N> 4) $ number of particles, the asymptotic behavior of the wave function in the complete configuration space remains insufficiently studied.

The most reliable means for finding the asymptotic behavior of wave functions in configuration space is the Fourier transform method, within which the wave function is given by the Fourier integral of the wave function in the momentum representation 
\cite {MerkFadd}. The asymptotic behavior of these integrals depends on the singularities of the wave functions in the momentum representation, which are described using the Lippmann-Schwinger integral equations (for a system of two particles) or the Faddeev-Yakubovsky equations (in the case of $ N \ge 3 $). It was this approach that made it possible for the first time to obtain correct asymptotic representations of wave functions for systems of three and four particles in the works \cite {Merkuriev1971, Yakovlev2016, Yakovlev1990}. It should be noted that the alternative approaches of \cite {Nuttal, Newton} did not allow obtaining the correct asymptotic behavior of the wave function for three particles. In this paper, we use the results obtained on the basis of the Yakubovsky integral equations in \cite {MerkFadd, MY1983, Yakovlev2014} to describe the structure of the wave function singularities of a system of $ N $ particles in the momentum representation for applying the Fourier transform method. It turned out that this method can be used to  study the weak asymptotics of $N$-particle wave functions with sufficient completeness. In this case, weak asymptotics are understood as asymptotic representations of wave functions considered as generalized functions (distributions) of angular variables as the radial variable tends to infinity. Such asymptotic formulas are of interest both from the point of view of the general theory of systems of several quantum particles, and from a practical point of view for setting asymptotic boundary conditions for the scattering problem in the hyperspherical representation. Let us clarify the concept of weak asymptotics using the well-known example of a system of two particles. The wave function 
$ \psi (x, k) $ defines a regular distribution by the integral
\begin{align}
\psi(|x|,k,g)=\int d{\hat x}\, \psi(|x|{\hat x},k)g(\hat x),  \label{I} 
\end{align}  
where $x,k \in \mathbb{R}^3$, ${\hat x}=x/|x| \in \mathbb{S}^2$, and functions  $g$ can be taken from a class of infinitely differentiable functions on   $\mathbb{S}^2$ .  Using asymptotics 
$$
\psi(x,k)\sim e^{i\langle k,x\rangle}+f(|k|{\hat x},k)|x|^{-1}e^{i|k||x|}
$$ 
as $|k||x|\to \infty$,  and appropriate integration by parts gives the asymptotic representation  
\begin{align}\label{TwoBodyPsi}
\psi
(|x|,k,g)\sim \frac{2\pi i}{|k|}\left\{\frac{e^{-i|k||x|}}{|x|}g(-\hat k)  - \frac{e^{i|k||x|}}{|x|} S(|k|,{\hat k};g) \right\} . 
\end{align}
Here and in what follows $i^2={-1}$, and the function 
\begin{align}
S(|k|,{\hat k};g)=g(\hat k) -\frac{|k|}{2\pi i}\int d{\hat x'}f(|k|{\hat x'},|k|{\hat k})g(\hat x') \nonumber 
\end{align}
is generated by the following singular kernel 
\begin{equation}
S(|k|,{\hat k_1},{\hat k_2})=\delta({\hat k_1},{\hat k_2})-\frac{|k|}{2\pi i}f(|k|{\hat k_1},|k|{\hat k_2}), 
\label{TwoBodyS}
\end{equation}
where $\delta({\hat k_1},{\hat k_2})$ is the delta function on a unit sphere  $\mathbb{S}^2$.  
This kernel
called the $s$-matrix, is the restriction of the kernel of the scattering operator \cite{MerkFadd} to the so-called energy surface. In the subsequent sections of the paper, a systematic generalization of these constructions is developed for the case of wave functions of a system of $N$ particles.

The paper has the following structure. Section 2 introduces the necessary notation and gives the problem statement. Here the Jacobi coordinates for $N$ particles and all the operators required in what follows are determined in sufficient detail, and the definitions of wave functions are also given. Section 3 describes the structure of the wave function singularities for a system of $N$ particles in the momentum representation. This is done in most detail for the case of three particles. Section 4 is the main one, where asymptotic representations of wave functions in the weak sense are derived. The fifth section contains applications of weak asymptotics to the hyperspherical representation of the scattering problem for $N$ particles.
In the conclusion, the main results of the paper are formulated.

\section{Formulation of the problem	}

This section is introductory, here we mainly follow the notations and definitions from 
\cite{MerkFadd, MY1983} and \cite{Yakovlev2014}.

A quantum system of $ N \ge 3 $  pairwise interacting nonrelativistic particles is considered. The nontrivial dynamics of the system is governed 
by the Hamiltonian of the relative motion of  particles. This operator is specified with the help of translation-invariant Jacobi coordinates. The system of such coordinates is not unique and its specific representatives are classified using the 
 partition chains \cite{Yakovlev2014}. Let us give the necessary definitions. Having numbered the particles with natural numbers from 1 to $ N $, we then simply identify the particle by its number. Partition $ a_k $ is a certain way of dividing a system of $ N $ particles into $ k $ subsystems:
$ a_k = \{\omega_1, \omega_2, ..., \omega_k \} $, $ \omega_i \subset \{1,2, ... N \} $, $ 
\omega_1 \bigcup \omega_2 \bigcup. .. \bigcup \omega_k = \{1,2, ..., N \} $, $ \omega_i \bigcap 
\omega_k = \emptyset $, $ i \ne k $. The order of the numbers of particles in the subsystems and the order of the subsystems are insignificant. Partition $ a_k $ follows 
$ a_l $ ($ a_k \subset a_l $) for $ k \ge l $ if: $ a_l = a_k $ for $ l = k $, for $ k = l + 1 $ for partitions $ a_l = \{\omega_1, ..., \omega_ {l-1}, \omega_l \} $ and $ a_ {l + 1} = \{ \omega_1, ..., 
\omega_ {l-1}, \omega'_ {l}, \omega' _ {l + 1} \} $ the relation $\omega'_l \bigcup \omega'_ {l + 1} = \omega_l $ is executed and for $ k> l + 1 $ there exists a sequence of successive partitions such that 
$ a_k \subset a_ {k-1} \subset ... \subset a_l $. In what follows, the same letters denote only the following one after another partitions connected by the $\subset $ sign.
A sequence of partitions of this type for $ k> l $ is called a partition chain
$ A^k_l = a_k, a_ {k-1}, ..., a_l $. It is clear that in the complete chain $ A^N_1 $ the extreme partitions are trivial therefore, it suffices to consider the chains $ A^k_l $ with  $ 2 \le l <k \le N-1 $. With these restrictions, the chain  $ A^{N-1} _2 $ is denoted as $ A_2 $. It is easy to calculate the total number $ {n} _N $ of such $ A_2 $ chains: $ n_N = 2^{1-N} N! (N-1)! $. Each chain $ A_2 $ is associated with a unique set of $ N-1 $ Jacobi vectors $ x^{a_2} _ {a_1}, x^{a_3 }_ {a_2}, ..., 
x^{a_l} _ {a_ {l -1}}, ..., x^{a_N} _ {a_ {N-1}} $ with $ x^{a_l} _ {a_ {l-1}} \in \mathbb{R}^3 $, which are constructed by the formulas	
\begin{equation*}
 x^{a_l}_{a_{l-1}}= \left(2\frac{M_{\omega_{l-1}} M_{\omega_l}}{M_{\omega_{l-1}}+ M_{\omega_l}}\right)^{1/2} \left(R_{\omega_{l-1}}-R_{\omega_l}\right).   
\end{equation*}
Here the subsystems $ \omega_ {l-1} $ and $ \omega_l $ are determined by the following partitions $ a_ {l-1} = \{\omega_1, ..., \omega_ {l-2}, \omega ' _ {l-1} \} $ and $ a_ {l} = \{\omega_1, ..., \omega_ {l-2}, \omega_ {l-1}, \omega_l \} $, such that $\omega'_ {l-1} = \omega_ {l-1} \bigcup 
\omega_l $. The masses $ M _ {\omega_k} $ of the  subsystems and the radius vectors of their centers of mass $ R_ {\omega_k} $ are given by the formulas
\begin{equation*}
M_{\omega_k}=\sum_{j\in\omega_k}m_j, \ \ R_{\omega_k}=M^{-1}_{\omega_k}\sum_{j\in\omega_k}m_jr_j
\end{equation*}    
in terms of masses $ m_j $ and radius vectors $ r_j $ of particles. A set of $ N-1 $ vectors 
$ x^{a_2} _ {a_1}, x^{a_3}_{a_2}, ..., x^{a_l}_{a_ {l-1}}, ... , x^{a_N}_{a_ {N-1}} $ defines a point 
$ X_{A_2} $ in $ \mathbb{R}^d $ with $ d = {3 (N-1)} $.
When it is required to emphasize that this is precisely the configuration space of the relative motion of $N$ particles, we denote it $\mathbb{R}^d_c $.
For further it is useful to group Jacobi vectors into two groups of vectors $ x_ {a_l} = 
\{x^{a_ {l + 1}}_{a_l} ... \, x^{a_N}_{a_{N-1} } \} \in \mathbb{R}^{3 (N-l)} $ being  internal with respect to the partition $ a_l $ and
	$ y_ {a_l} = \{x^{a_2}_{a_1}, x^{a_3}_{a_2}, ..., x^{a_l}_{a_{l-1}} \} \in \mathbb{R}^{3 (l-1)} $  being  external with respect to the partition $ a_l $ and such that $ X_ {A_2} = \{x_{a_l}, y_{a_l} \} $.
In turn, the vectors forming $ x_ {a_l} $ can also be divided into two groups
$ x_{a_i} = \{x^{a_{i + 1}}_{a_i} ... \, x^{a_N}_{a_{N-1}} \} $ being  internal with respect to the partition 
$ a_i \subset a_l $ and
	$ y_{a_i a_l} = \{x^{a_{l + 1}}_{a_l}, x^{a_{l + 2}}_{a_{l + 1}}, ..., x^{a_i}_{a_ {i-1}} \} $  being external with respect to the partition $ a_i $, but internal with respect to the partition $ a_l $.
		The introduced coordinates allow for a more detailed grouping of Jacobi vectors with respect to the pair of partitions $ a_i \subset a_l $, such that 
		$ x_{a_l} = \{x_{a_i}, y_{a_i a_l} \} $, and $ y_{a_i} = \{y_{a_ia_l}, y_{a_l} \} $.
 
The Hamiltonian $ \bf H $ of the relative motion of a system of $ N $ particles in the coordinate representation acts as an operator in $ L_2(\mathbb {R}^d_c) $ and is given by the expression
\begin{equation*}\label{H}
{\bf H}=
-\Delta_{X_{A_2}}
+ \sum_{b_{N-1}} V_{b_{N-1}}(x_{b_{N-1}}), 
\end{equation*}
where $ \Delta_{X_{A_2}} $ denotes the Laplacian in $ \mathbb{R}^d_c $, equal to the sum of the three-dimensional Laplace operators by the components of $ X_{A_2} $ Jacobi vectors.
The representation $ \Delta_{X_{A_2}} = \Delta_{x_{a_l}} + \Delta_{y_{a_l}} $ is also true for any 
$ a_l $ with $ 2 \le l \le N-1 $. In \cite{Yakovlev2014} it is shown that the transition from coordinates $ X_{A_2} $ to coordinates $ X_{B_2} $ at $ A_2 \ne B_2 $ is carried out using an orthogonal transformation, which leads to 
$ \Delta_{X_{A_2} } = \Delta_{x_{a_l}} + \Delta_{y_{a_l}} = \Delta_{X_{B_2}} = \Delta_{x_{b_{l}}} 
+ \Delta_{y_{b_l}} $ and for this reason we will henceforth omit the chain index in the notation of vectors in $ \mathbb {R}^d_c $ and write simply $X$. The second sum with interaction potentials $ V_{b_{N-1}} $ in (\ref {H}) extends to all possible partitions 
$b_ {N-1}$ and consists of 
$N(N-1) / 2$ terms.  Since the partition $ b_{N-1} $ contains only one nontrivial subsystem of two particles with numbers, say, $i, j$, then each term $V_{b_{N-1}} (x)$ corresponds to the interaction potential of the pair of particles $i, j$. It is assumed that 
$ V_{b_{N-1}}(x) $ are sufficiently smooth and sufficiently rapidly decreasing functions 
$$
V_{b_{N-1}}(x) \sim O(|x|^{-\beta}), \ \ \beta>1
$$
for $|x| \to \infty$  with $ \beta $ such that the requirements of the work \cite {Faddeev1963} are satisfied.

Together with the operator $ \bf H $ we also consider  the kinetic energy operator
$$
{\bf H}_0=-\Delta_X
$$    
and operators
$$
{\bf H}_{a_l}={\bf H}_0+{\bf V}_{a_l} \equiv {\bf H}_0+\sum_{a_{N-1}\subset {a_l}}
{\bf V}_{a_{N-1}}, 
$$
in which the interaction potentials act only between particles belonging to subsystems of   the partition $ {a_l} $. For resolvents of the introduced operators, the standard notations are used
$$
{\bf R}(z)=({\bf H}-z)^{-1},\ \ {\bf R}_0(z)=({\bf H}_0-z)^{-1}, \ \ {\bf R}_{a_l}=({\bf H}_{a_l}-z)^{-1}. 
$$ 
We also define the $ T $-matrices closely related to resolvents by 
\begin{equation}\label{T-R}
	{\bf T}_{a_l}(z)= {\bf V}_{a_l}-{\bf V}_{a_l}{\bf R}_{a_l}(z){\bf V}_{a_l}. 
\end{equation}
For $ l = 1 $ the formula (\ref {T-R}) becomes
\begin{equation}
{\bf T}(z)= {\bf V}-{\bf VR}(z){\bf V},   \ \label{T}
\end{equation}
where $ {\bf V} = \sum_ {a_{N-1}} {\bf V}_{a_{N-1}} $, and it gives the complete 
$ N $-particle $ T $-matrix. In terms of $ T $-matrices, the resolvents are reconstructed by the formulas
\begin{equation}\label{R-T}
{\bf R}_{a_l}(z)={\bf R}_0(z)-{\bf R}_0(z){\bf T}_{a_l}(z){\bf R}_0(z). 
\end{equation}

The wave functions of a system of $ N $ particles in the continuum studied in this work are determined using the resolvent by the following expression \cite {MerkFadd, MY1983, Yakovlev2014}
\begin{equation}\label{Psi}
\Psi^{\pm}(P) = \lim_{\epsilon\to +0}(\mp i\epsilon) {\bf R}(P^2\pm i\epsilon)\Phi_0(P). 
\end{equation}
Here $ P \in \mathbb{R}^d $, and $ \Phi_0 (P) $ is an eigenfunction of the operator 
$ {\bf H}_0 $
$$
{\bf H}_0 \Phi_0 (P) = P^2 \Phi_0 (P),
$$
which is given by the plane wave $ \Phi_0 (X, P) = (2 \pi)^{- d / 2} 
\exp \{i \langle P, X \rangle \} $ in coordinate representation. Here and in what follows, the symbol $ \langle P, X \rangle $ denotes the scalar product of the vectors $ P $ and $ X $.
	By construction, $ \Psi^\pm (P) $ satisfies the equation
$$
{\bf H}\Psi^\pm(P)=P^2\Psi^\pm(P). 
$$ 
Using (\ref{R-T}), the limit in (\ref {Psi}) is calculated, leading  to the following result
\begin{equation}
\Psi^\pm(P)=\Phi_0(P)-{\bf R}_0(P^2\pm i0){\bf T}(P^2\pm i0)\Phi_0(P). 
\label{Psi-pm}
\end{equation}  

In the coordinate representation, the wave functions $ \Psi^\pm (X, P) $ depend on the coordinate variable $ X \in \mathbb {R}^d $ and on the conjugate momentum variable 
$P \in \mathbb {R}^d $. These functions define regular distributions using the integrals
\begin{align}
\Psi^\pm(|X|,P;G)=\int d {\hat X}\,\Psi^\pm(|X|{\hat X},P)G
({\hat X}) \label{PsiX}, 
\end{align}  
where $ | X | $ 
is the modulus of the vector $ X $, which is also called the hyperradius,
$ {\hat X} = X / | X | \in \mathbb{S}^{d-1} $  is 
the unit vector in the direction of $ X $, and the trial functions $ G $ 
can be taken, for example, from the class of infinitely differentiable functions on the unit sphere $ \mathbb{S}^{d-1} $.

The distributions (\ref{PsiX}) are a generalization of (\ref{I}) to the case of $ N $ particles, and the task of the next sections is to obtain the asymptotics of the functions 
$ \Psi^\pm(| X |, P; G) $ 
as $ | X | \to \infty $, which will be a generalization of the formula (\ref{TwoBodyPsi}) to the case of $ N $ particles.
In fact, in what follows, only functions with the $ + $ sign are considered, since the behavior of functions with the $ - $ sign, with few exceptions, repeats the behavior of functions with the $ + $ sign and can be found by similar tricks. For this reason, in what follows, the $ + $ sign in the notation of wave functions is omitted.

In the next sections we need wave functions and binding energies of bound states of particles belonged to subsystems of the partition  $ a_l $. 
These functions are defined as square-integrable solutions  of the equation 
\begin{equation*}
	[-\Delta_{x_{a_l}}+V_{a_l}(x_{a_l})  ] \psi_{a_l}(x_{a_l})=-\lambda^2_{a_l}\psi_{a_l}(x_{a_l}).
	\label{psi_a_l}
\end{equation*}
For the functions $ \psi_{a_l} $ and the binding energies $ - \lambda^2_{a_l} $, here and in what follows, it is assumed that the index $ a_l $ also implicitly includes all the necessary quantum numbers that identify these eigenfunctions and the corresponding eigenvalues.  For the latter, it is assumed that $ \lambda^2_{a_l}> 0 $. In addition to the functions 
$ \psi_{a_l} $,  we will also need form factors defined with their help by 
\begin{equation}
\phi_{a_l}=-{\bf V}_{a_l}\psi_{a_l}.
\label{phi_a_l}
\end{equation}

\section{Wave function structure in momentum \\ representation}
\subsection{General case of $ N $ particles}
Momentum space $ \mathbb {R}^d_m $ is 
formed by vectors $ P \in \mathbb{R}^d$, which are conjugate to Jacobi vectors $ X $
from $ \mathbb{R}^d_c $.
Similar to the representation
$ X = \{x_ {a_l}, y_ {a_l} \} $  momentum  variables are also grouped 
$ P = \{k_{a_l}, p_{a_l} \} $ into internal $ k_{a_l} \in \mathbb {R} ^ {3 (N-l)} $ momentum and outer 
$ p_{a_l} \in \mathbb {R}^{3 (l-1)} $  momentum with respect to the partition $ a_l $.
Relationship between functions in momentum $ {\hat f} (P) $ and coordinate $ f (X) $ representations carried out by means of the Fourier transform
$$
f(X)=\int_{\mathbb{R}^d_m} dP\, \Phi_0(X,P){\hat f}(P)
$$
with $\Phi_0(X,P)=(2\pi)^{-d/2}\exp\{i\langle P,X \rangle    \}$. 

In momentum representation the wave function ${\hat \Psi}(P',P)$  according to (\ref{Psi-pm})  is given by the formula 
\begin{equation}
{\hat \Psi}(P',P)=\delta(P'-P)-\frac{T(P',P,P^2+i0)}{P'^2-P^2-i0},
\label{Psi-PP}
\end{equation}  
where $P',P \in \mathbb{R}^d_m$, $\delta(P'-P)$  is delta function and $T(P',P,P^2+i0)$ is the kernel of the operator ${\bf T}(z)$ from  (\ref{T}). 
Note that in the case of two particles, the kernel $ T(P', P, P^2 + i0) $ is a smooth function of its arguments and, therefore, all singularities of $ {\hat \Psi} (P', P) $ are represented explicitly by singularities displayed  in (\ref {Psi-PP}): the delta function and the denominator in the second term of the right-hand side of (\ref {Psi-PP}). In the case of $ N \ge 3 $, this kernel contains several types of additional delta-functional and pole-wise singularities 
\cite {MerkFadd}. To describe the strongest delta-functional singularities, it is convenient to introduce the concept of a connected in the partition $ a_l $ operator 
\cite {MerkFadd, Yakovlev2014}: an integral operator with a kernel of the form
$ w_{a_l} (k'_{a_l}, k_{a_l}) \delta (p'_{a_l} -p_{a_l}) $, where $ w_{a_l} (k'_{a_l}, k_{a_l}) $ has no delta-functional singularities, is called connected in the partition $ a_l $ and is denoted by $ {\bf W}^c_{a_l} $. In the case $ l = 1 $, the corresponding operator is simply called connected.

All operators $ {\bf T}_{a_l} (z) $ for $ l <N-1 $ defined in (\ref {T-R}) are not connected. The connected parts of the operators $ {\bf T}_{a_l} (z) $ are constructed using the subtraction procedure defined by the following lemma, the proof of which can be found in 
\cite {MerkFadd} and \cite{Yakovlev2014} (we also give a proof of this Lemma below for the case $ N = 3 $): \\
{\bf Lemma 1.} {\it The connected parts of the operators $ {\bf T}_{a_l} (z) $, 
	$ 1 \le l \le N-1 $, are given by the formulas
\begin{equation}
{\bf T}^c_{a_l}(z)= {\bf T}_{a_l}(z)-\sum_{i=l+1}^{N-1}\sum_{a_i\subset a_l}{\bf T}^c_{a_i}(z),  
\label{T^c}
\end{equation}
where  ${\bf T}^c_{a_{N-1}}(z)={\bf T}_{a_{N-1}}(z)$.} \\ 
Since the kernel of the operator $ {\bf T}_{a_ {N-1}} (z) $, given by the expression
\begin{equation*}
t_{a_{N-1}}(k'_{a_{N-1}}, k_{a_{N-1}},z-p^2_{a_{N-1}})\delta(p'_{a_{N-1}}-p_{a_{N-1}}), 
\label{t-N-1}
\end{equation*}
by construction is connected in $ a_{N-1} $, then the formula (\ref{T^c}), starting with 
$ l = N-2 $, allows recursively constructing connected parts for all operators
$ {\bf T}_{a_l} (z) $. At the same time, the formula (\ref{T^c}) gives the required representation
\begin{equation}
{\bf T}_{a_l}(z)= \sum_{i=l+1}^{N-1}\sum_{a_i\subset a_l}{\bf T}^c_{a_i}(z) + {\bf T}^c_{a_l}(z)
\label{TT}
\end{equation} 
for  ${\bf T}_{a_l}(z)$ in terms of connected parts of operators for partitions which are followed the partition  $a_l$.  
In what follows, we need a formula (\ref{TT}) in terms of the kernels of the operators involved in it:
\begin{align}
T_{a_l}(P',P,z)&= t_{a_l}(k'_{a_l},k_{a_l},z-p^2_{a_l})\delta(p'_{a_l}-p_{a_l}), \label{Tal-k}\\  
t_{a_l}(k'_{a_l},k_{a_l},z)&= 
 \sum_{i=l+1}^{N-1}
 \sum_{a_i\subset a_l}
 t^c_{a_i}(k'_{a_i},k_{a_i},z-p^2_{a_i a_l})\delta(p'_{a_i a_l}-p_{a_i a_l})
 + t^c_{a_l}(k'_{a_l},k_{a_l},z), 
 \label{tal-k}
 \end{align} 
where $p_{a_i a_l}$ is the momentum conjugate to the coordinate  $y_{a_i a_l}$.  
In turn, the kernel 
$ t_{a_i} (k'_{a_i}, k_{a_i}, z) $ have two types of pole singularities, the explicit form of which can be obtained using the Faddeev-Yakubovsky equations \cite {MerkFadd, Yakovlev2014}. The first type includes singularities generated by bound states of particles included in the subsystems of the $ a_i $ partition. These singularities  are reflected in the  representation 
\begin{align}
t^c_{a_i}(k'_{a_i},k_{a_i},z) &= \frac{\phi_{a_i}(k'_{a_i}){\bar \phi_{a_i}}(k_{a_i})}
{z+\lambda^2_{a_i} }+ \nonumber \\ 
\sum_{j,m > j} \   \sum_{a_j,a_m\subset  a_i} 
&\frac{\phi_{a_j}(k'_{a_j})}{z +\lambda^2_{a_j}   - p'^2_{a_j a_i}  }
{\cal H}^{a_i}_{a_j a_m}(p'_{a_j a_i},p_{a_m a_i},z)
\frac{{\bar \phi_{a_m}}(k_{a_m})}{z +\lambda^2_{a_m}- p^2_{a_m a_j}}, 
\label{tc}
\end{align} 
where in summation the possibility  $j, m=N$ is permitted and in this case it should be assumed 
$$
\frac{\phi_{a_N}(k_{a_i})}{z +\lambda^2_{a_N}    - p^2_{a_N a_i}   }\equiv 1. 
$$
Hereinafter, the bar above the function symbol means complex conjugation.
The form factors $ \phi_{a_j} (k_{a_j}) $ in the formula (\ref {tc}) are defined as Fourier transforms of functions (\ref{phi_a_l})
$$
\phi_{a_j}(k_{a_j})=
(2\pi)^{-3(N-j)/2}
\int dx_{a_j}\,e^{-i\langle k_{a_j},x_{a_j}\rangle} \phi_{a_j}(x_{a_j}). 
$$
The second type singularities of the kernels $ t^c_{a_i} (k '_ {a_i}, k_ {a_i}, z) $ is caused by the processes of successive rescattering of individual particles and particles bound into clusters included in subsystems of the partition  $ a_i $ 
\cite{Merkuriev1971,Yakovlev2016,Yakovlev1990,MerkFadd}. 
In general, these singularities have the form of products of poles and are described by rather cumbersome formulas, as, for example, it was demonstrated in the works 
\cite {Yakovlev1990, MerkFadd} for systems of three and four particles. We give below the corresponding formulas for the case of three particles,
however, the characteristic properties of these singularities will remain valid in the general case.

\subsection{Case $N=3$}
For $ N = 3 $, only partitions of the type $ a_2 $ into two subsystems are nontrivial, consisting of a pair of particles in one subsystem and a third particle: $ a_2 = \{(ij) k \} $, and the chains $ A_2 $ consist only of one partition of this type. In order to reduce the cumbersomeness of the formulas, we will drop the index "2"\   in the notation of partitions for the considered case of a system of three particles.
The $ T $-matrix of a system of three particles is given by the sum of the components
\begin{equation*}\label{T3}
{\bf T}(z)=\sum_{ab}{\bf M}_{ab}(z), 
\end{equation*}
which, in turn, obey the Faddeev system of equations \cite{Faddeev1963} 
\begin{equation}\label{IFE}
{\bf M}_{ab}(z)={\bf T}_a(z)\delta_{ab}-{\bf T}_{a}(z){\bf R}_0(z)\sum_{c\ne a} {\bf M}_{cb}(z), 
\end{equation}
where  $\delta_{ab}$  is the Konecker symbol. After one ineration the equations (\ref{IFE}) take the form 
\begin{align}
{\bf M}_{ab}(z)={\bf T}_a(z)\delta_{ab}-{\bf T}_{a}(z){\bf R}_0(z) {\bf T}_b(z)(1-\delta_{ab}) \nonumber \\
+
{\bf T}_a(z){\bf R}_0(z)\sum_{c\ne a}{\bf T}_c(z){\bf R}_0(z) \sum_{f\ne c} {\bf M}_{fb}(z).
\label{IFE-iter}
\end{align}
As mentioned before, the operator ${\bf T}_a(z)$  is connected in the partition $a$  and has  the kernel 
\begin{equation*}
t_a(k'_a,k_a,z-p^2_a)\delta(p'_a-p_a).
\label{t_a}
\end{equation*}
The kernel of the operator ${\bf T}_a(z){\bf R}_0(z){\bf T}_b(z)$  as  $a\ne b$ is calculated with the help of the delta-functions which are included in the kernels of $T$-matrices and has the form 
\begin{equation}
\frac{t_a(k'_a,k_a(p_b,p'_a),z-p'^2_a)t_b(k_b(p'_a,p_b),k_b,z-p^2_b)}
{|s^3_{ba}|[k^2_b(p'_a,p_b)+p^2_b-z]}.
\label{tatb}
\end{equation}
In calculating the integrals for the kernels of the operators $ {\bf T}_a (z) {\bf R}_0 (z) 
{\bf T}_b (z) $, we used the following transformation formulas for the relative momenta
 \cite{MerkFadd} 
\begin{align}
k_b&=c_{ba}k_a+s_{ba}p_a, \nonumber \\
p_b&=-s_{ba}k_a+c_{ba}p_a, 
\label{ba}
\end{align}
where the coefficients $ c_{ba} $ and $ s_{ba} $ depending only on the masses of particles satisfy the relation $ c^2_{ba} + s^2_{ba} = 1 $, which ensures the orthogonality of the transformations (\ref {ba}). The representation for $ k_a $ obtained from the second equality in (\ref{ba}) determines the value of $ k_a (p_b, p_a) $
\begin{equation}
k_a(p_b,p_a)=-\frac{1}{s_{ba}}p_b+\frac{c_{ba}}{s_{ba}}p_a, 
\label{kpp}
\end{equation}
 used in $(\ref{tatb})$.  The expression for $k_b(p_a,p_b)$ is obtained form  
 (\ref{kpp}) by an appropriate  interchange of indices. 

According to (\ref {tatb}), the operator $ {\bf T}_a (z) {\bf R} _0 (z) {\bf T}_b (z) $ is connected and the same is true for the same operator with $ b = c $ in the kernels of operators in the last group of terms in (\ref {IFE-iter}). This means that the third group of summands on the right-hand side of (\ref{IFE-iter}) contains only connected operators.
The obtained representations prove the assertion of Lemma 1 and, accordingly, the formula (\ref {TT}) for the case of three particles. Although the kernel (\ref {tatb}) does not contain 
$ \delta $-functional singularities, however, it remains singular due to the presence of the  denominator, which can vanish for non-negative real $ z $.
In the paper \cite {Faddeev1963}, it was shown that the third term in (\ref {IFE-iter}), being a connected operator, no longer has pole singularities similar to (\ref {tatb}), but may contain only weak logarithmic singularities.


We have actually described the above-mentioned second type of singularities of the 
$ T $-matrix kernel for a system of three particles. Let us move on to the description of the singularities identified in the formula (\ref {tc}) for the general case, which appear due to the presence of singular denominators in two-particle $ t $-matrices \cite {Faddeev1963}
 \begin{equation*}
 t_a(k'_a,k_a,z)=\frac{\phi_a(k'_a) \overline{\phi}(k_a)}{z+\lambda^2_a}+{\hat t}_a(k'_a,k_a,z).
 \label{ta}
 \end{equation*} 
 Here   ${\hat t}_a(k'_a,k_a,z)$  is a smooth function of its arguments. Let us calculate the numerator in (\ref {tatb}): 
 \begin{align}
 {\hat t}_a(k'_a,k_a(p_b,p'_a),z-p'^2_a){\hat t}_b(k_b(p'_a,p_b),k_b,z-p^2_b)+      \nonumber    \\ 
 \frac{\phi_a(k'_a)}{z+\lambda^2_a-p'^2_a}\overline{\phi}_a(k_a(p_b,p'_a)) 
 {\hat t}_b(k_b(p'_a,p_b),k_b,z-p^2_b)+ \nonumber \\
 {\hat t}_a(k'_a,k_a(p_b,p'_a),z-p'^2_a) \phi_b(k_b(p'_a,p_b)) \frac{\overline{\phi}_b(k_b)}{z+\lambda^2_b-p^2_b} + \nonumber \\
 \frac{\phi_a(k'_a)}{z+\lambda^2_a-p'^2_a}\overline{\phi}_a(k_a(p_b,p'_a))\phi_b(k_b(p'_a,p_b)) \frac{\overline{\phi}_b(k_b)}{z+\lambda^2_b-p^2_b}. 
\label{tatb-sing}
 \end{align}
 It is important to note that the zeros of the denominators in (\ref{tatb-sing}) and the zeros of the denominator in (\ref {tatb})  do not intersect
 \begin{align}
 k^2_b(p'_a,p_b)+p^2_b-z + z+\lambda^2_b-p^2_b = k^2_b(p'_a,p_b) + \lambda^2_b\ge  \lambda^2_b > 0 ,\nonumber \\
 k^2_b(p'_a,p_b)+p^2_b-z=k^2_a(p_b,p'_a)+p'^2_a-z ,\nonumber \\ 
 k^2_a(p_b,p'_a)+p'^2_a-z + z+\lambda^2_a-p'^2_a = k^2_a(p_b,p'_a) + \lambda^2_a\ge  \lambda^2_a > 0. 
 \label{denomin}
 \end{align} 
 For this reason, the corresponding singularities can be considered separately.
 From the structure of the equations (\ref {IFE-iter}) it follows that the kernel of the last term will have a similar to (\ref {tatb-sing}) form. As a result, for the kernel of the operator
$$
{\bf W}_{ab}(z)={\bf M}_{ab}(z)-{\bf T}_a(z)\delta_{ab}
$$  
one obtains the representation  \cite{Faddeev1963}
\begin{align}
&W_{ab}(P',P,z)= F_{ab}(P',P,z)   +   \frac{\phi_a(k'_a)}{z+\lambda^2_a-p'^2_a} G_{ab}(p'_a,P,z) + \nonumber \\
&J_{ab}(P',p_b,z)\frac{\overline{\phi}_b(k_b)}{z+\lambda^2_b-p^2_b}  + 
\frac{\phi_a(k'_a)}{z+\lambda^2_a-p'^2_a}H_{ab}(p'_a,p_b,z) \frac{\overline{\phi}_b(k_b)}{z+\lambda^2_b-p^2_b}.
\label{FGJH}
\end{align} 
Thus, we have described the structure of all  singularities of the three-particle 
$T$-matrix, which also determine all singularities of the three-particle wave function 
$ {\hat \Psi} (P', P) $,
expressed in terms of the $ T $-matrix by the formula (\ref {Psi-PP}).
 
\section{Weak asymptotics of the wave function of a system of $ N $ particles}
\subsection{Asymptotic representations}
In this section, we obtain the asymptotic for $ | X | \to \infty $
\footnote {In fact, a large parameter in calculating the asymptotics of wave functions is the product $ | P || X | $, however, we assume in what follows that the value
of	$|P|$ is bounded and nonzero, so the requirement $ | X | \to \infty $ is enough. }
representation for distribution (\ref {PsiX})
\begin{equation}
	\Psi(|X|,P;G)=\int_{\mathbb{S}^{d-1}} d{\hat X}\, \Psi(X,P)G({\hat X}), 
\label{Psi-d}
\end{equation} 
called the weak asymptotics of the wave function $\Psi(X,P)$. 
Functions
$ G (\hat X) $ are taken from the set of infinitely differentiable functions on the unit sphere $ {\hat X} \in \mathbb{S}^{d-1} $ in the space $ \mathbb {R}^d $. Using the expression for 
$ \Psi (X, P) $ in terms of the wave function in the momentum representation
\begin{equation*}
\Psi(X,P)=\int_{\mathbb{R}^d} dP'\, \Phi_0(X,P'){\hat \Psi}(P',P) 
\label{Psi-F}
\end{equation*}
and the formula  (\ref{Psi-PP}), we obtain the representation for the integral  (\ref{Psi-d}) 
\begin{align}
\Psi(|X|,P;G)&=\int_{\mathbb{S}^{d-1}} d{\hat X}\, \Phi_0(X,P)G({\hat X}) - \nonumber \\
&- \int_{\mathbb{R}^d} dP'\, \frac{T(P',P,P^2+i0)}{P'^2-P^2-i0}\int_{\mathbb{S}^{d-1}} d{\hat X}\, \Phi_0(X,P')G({\hat X}).
\label{Psi-dd}
\end{align}
In the right hand side of (\ref{Psi-dd})  it is appeared an universal integral of the form 
\begin{equation}
J(|X|,P)=\int_{\mathbb{S}^{d-1}} d{\hat X}\, \Phi_0(X,P)G({\hat X}),
\label{J-univ}
\end{equation}  
 where $\Phi_0(X,P)= (2\pi)^{-d/2}\exp\{i\langle P,X\rangle\} $.  
 Asymptotics of this integral as $|X| \to \infty$ is given by the following lemma: 
\\ 
{\bf Lemma 2.} {\it The integral} $J (| X |, P)$ {\it for} $ | P || X | \to \infty $ {\it has an asymptotic representation}  \\ 
\begin{align*}
J(|X|,P) \sim N_d(P)
\left[Q_d^-(|P|,|X|)
G({-\hat P}) - Q_d^+(|P|,|X|) 
G({\hat P})             
\right], 
\end{align*}
{\it where the spherical waves $Q_d^\pm(|P|,|X|)$ in $\mathbb{R}^d$ and the coefficient  $N_d(|P|) $  are determined by formulas:	}
	$$ 
	Q_d^\pm(|P|,|X|)=
	\frac {\exp\{\pm i|P||X|  \mp i\pi(d-3)/4\}}{|X|^{\frac{d-1}{2}}}, \ \ N_d(|P|)=
	i   (2\pi)^{-\frac{1}{2}}|P|^{-\frac{d-1}{2}}.
	$$
Although the proof of this statement can be found in the literature, for completeness we give it in the second part of this section.

Lemma 2 solves the problem of the asymptotic behavior of the first term in 
(\ref {Psi-dd}). To construct the asymptotics of the second term in this formula, it is necessary to calculate the integral over the variable $ P'$. This integral, using Lemma 2, asymptotically takes the form of the difference
\begin{align*}
I^{-}(|X|,P)-I^+(|X|,P), 
\label{II}
\end{align*}
where 
\begin{equation}
I^{\pm}(|X|,P)= \int_0^\infty |P'|^{d-1}d |P'| \frac{Q^+_d(|P'|,|X|)}{P'^2-P^2-i0} N_d(|P'|)F^{\pm}(|P'|,P;G).
\label{Ipm}
\end{equation}
The quantities  $F^\pm$ are expressed in terms of the $T$-matrix kernel   by the formulas 
\begin{align}
F^\pm(|P'|,P;G)= \int_{\mathbb{S}^{d-1}} d{\hat P'}\, T(|P'|{\hat P'},P,P^2+i0)G(\pm {\hat P'}) 
\label{Fpm}
\end{align} 
and, as it is proved in the second part of this section,  they are sufficiently smooth functions.   
In this case, the asymptotics of the one-dimensional integrals $ I^\pm(| X |, P) $ from 
(\ref {Ipm}) as $ | X | \to \infty $ is obtained using an appropriate closure of the integration contours in the upper (+) or the lower (-) complex half-plane and the residue theorem.   For the integral $ I^+ $ the asymptotics has the form 
\begin{equation*}
I^+(|X|,|P|)\sim i \pi|P|^{d-2} Q^+_d(|P|,|X|)N_d(|P|)F^+(|P|,P;G). 
\label{I+as}
\end{equation*}
The integral $ I^- $ in this case has an exponentially small order for $ | X | \to \infty $.
Collecting the contributions obtained, we arrive at the final asymptotic representation for
$ \Psi (| X |, P; G) $ in the form
\begin{align}
\Psi(|X|,P;G) \sim N_d(|P|)
\left\{  Q^{-}_d(|P|,|X|)G(-{\hat P}) -  Q^{+}_d(|P|,|X|)S(|P|,{\hat P};G) \right\}   , 
\label{NBodyPsiWasymp} 
\end{align}
where 
\begin{align}
S(|P|,{\hat P};G)= G({\hat P}) - i\pi |P|^{d-2} F^+(|P|,P;G) .  
\label{SS-N}
\end{align}
Note the similarity of the structures of the obtained formulas (\ref {NBodyPsiWasymp}, 
\ref {SS-N}) and the asymptotics
(\ref {TwoBodyPsi}). As in the two-particle case, the function $ S (| P |, {\hat P}, G) $ is generated by the formal singular kernel

\begin{equation}
S(|P|,{\hat P}_1,{\hat P}_2) = \delta({\hat P}_1,{\hat P}_2)-i\pi |P|^{d-2} T(|P|{\hat P}_1,|P|{\hat P}_2,P^2+i0), 
\label{S-N}
\end{equation}
where $ \delta ({\hat P}_1, {\hat P} _2) $ is  delta function on the unit sphere 
$ \mathbb{S}^{d-1} $. The kernel (\ref {S-N}) defines an integral operator acting according to the formula (\ref {SS-N}) on functions defined on the unit sphere 
 $\mathbb{S}^{d-1}$. 
As in the two-particle case, this kernel is related to the scattering operator \cite {MerkFadd} of the $ N $ particle system. However, as we show in Section 5, in contrast to the case of two particles, it is a restriction on the energy surface of only one of the matrix elements
of the $ N $ particles scattering operator corresponding to transitions between scattering channels in which all $ N $ particles are free \cite {MerkFadd, Faddeev1963}. This kind of phenomenon is usually called filtration. In this context, asymptotic filtering occurs, as a result of which the leading terms of the weak asymptotics of the wave function contain only the contribution from scattering processes in which all $ N $ particles are free both before and after the interaction. The contribution from processes with reactions of formation of bound states of subsystems of particles and successive rescattering of particles in the leading terms of the weak asymptotics is absent.

The formula (\ref {NBodyPsiWasymp}) is the main result of this section and defines
weak asymptotics of the wave function of a system of $ N $ particles as the superposition of incoming and outgoing  spherical waves, in which the coefficient of the outgoing  spherical wave is expressed in terms of the $ S $-matrix for $ N $ particles. Note that for $ N = 2 $ the formulas (\ref {NBodyPsiWasymp} - \ref {S-N}) reproduce the two-particle result 
(\ref {TwoBodyPsi}, \ref {TwoBodyS}), in which the scattering amplitude is expressed in terms of $ t $-matrix by the well-known formula \cite {MerkFadd}
$$
f(|k|{\hat k}_1,|k|{\hat k}_2)=-2\pi^2 t(|k|{\hat k}_1,|k|{\hat k}_2,k^2+i0).
$$
 In the second part of this section, we give proofs of the key facts used to derive the asymptotics (\ref{NBodyPsiWasymp}): Lemma 2 and the smoothness properties of the functions $ F^ \pm $.

\subsection{Proofs}
To prove Lemma 2, consider the integral
\begin{equation}
J(p,|x|)= \int_{\mathbb{S}^{d-1}} d{\hat x} \, \exp\{i\langle p,x\rangle \} g({\hat x}), 
\label{J}
\end{equation}
where $x,p \in \mathbb{R}^d$, ${\hat x}=x/|x| \in \mathbb{S}^{d-1}$  and $|x|\to \infty$. 
Let us convert the scalar product in the exponent to the form
$$
\langle p,x \rangle = \langle x',q \rangle, \ \ x'=|x|{\hat p}, \ \ q=|p|{\hat x}   
$$
and rewrite the integral (\ref {J}) in new variables, complementing it to a complete integral over $ \mathbb {R}^d $ using  delta function
\begin{equation}
J(p,|x|)= 2\int_{\mathbb{R}^d}dq\, |q|^{-(d-2)}\delta(q^2-p^2)\exp\{i\langle x',q \rangle\} g({\hat q})h(q^2).
\label{J-delta}
\end{equation}
Here we have introduced an auxiliary smooth compactly supported function $ h $ such that $ h (p^2) = 1 $. Let us further use the well-known representation for the delta function
$$
2\pi i \, \delta(q^2-p^2)= (q^2-p^2-i0)^{-1}-(q^2-p^2+i0)^{-1}
$$
and transform (\ref {J-delta}) to the difference of two integrals 
$$
J(p,|x|) = J^+(p,|x|)-J^-(p,|x|), 
$$
where 
$$
J^{\pm}(p,|x|)=  
\frac{1}{2\pi i}\int_{\mathbb{R}^d}dq\, 
\frac{\exp\{i\langle x',q \rangle\}}{q^2-p^2 \mp i0} 
g({\hat q})h(q^2)|q|^{-(d-2)}.
$$
The asymptotics of integrals of the form $ J^\pm (p, | x |) $ as $ | x | \to \infty $ is calculated using the well-known asymptotic representation, in particular, which was used many times in the papers \cite{Merkuriev1971}-\cite{MerkFadd}, 
\begin{align}
\int_{\mathbb{R}^d}  dq \, \frac{\exp\{i\langle x,q \rangle\}}{q^2-p^2 \mp i0} f(q) \sim  &\pi (2\pi)^{(d-1)/2}|p|^{(d-3)/2} f(\pm |p|{\hat x}) \times     \nonumber \\ 
&\times \frac{\exp\{\pm i |p||x| \mp i\pi(d-3)/4   \}}{   |x|^{(d-1)/2}}. 
\label{as-int-pol}
\end{align}
Applying  (\ref{as-int-pol}) to the integrals $J^\pm$, we obtain the final result for the asymptotics of the integral  $J(p,|x|)$  
\begin{align}
&J(p,|x|)\sim  i \left(\frac{2\pi}{|p|}\right)^{(d-1)/2} \times  \label{Jas}
\\ 
&\times 
\left[
\frac{\exp\{-i|p||x|+i\pi (d-3)/4\} }{|x|^{(d-1)/2}}g(-{\hat p})
-\frac{\exp\{i|p||x|-i\pi (d-3)/4\}}{|x|^{(d-1)/2}}  g({\hat p})
 \right]. \nonumber
\end{align} 
Note that, as expected, the auxiliary function $ h $ did not contribute to (\ref {Jas}) due to the fact that
$ h (p^2) = 1 $.
The statement of Lemma 2 is now obtained by applying (\ref {Jas}) to the integral 
(\ref {J-univ}).

Let us turn to the study of the smoothness properties of the functions $ F^{\pm} $ defined by integrals over the angular variables
(\ref {Fpm}). As shown in Section 3.1, the kernel of the $ T $-matrix included in (\ref {Fpm}) has delta-function singularities. This kernel is given by formulas (\ref {Tal-k}) and 
(\ref {tal-k}) for $ l = 1 $ and, therefore, the integrals in (\ref {Fpm}) become the sum of integrals over angular variables with delta functions in integrands. A typical term in these sums has the form
\begin{equation}
D_m(|P'|,P)= \int_{\mathbb{S}^{d-1}} d{\hat P'} \, t^c_{a_m}(k'_{a_m},k_{a_m},k^2_{a_m}+i0)\delta(p'_{a_m}-p_{a_m})G({\hat P'}).
\label{D-m}
\end{equation}        
To calculate an integral of this kind, we will use the already familiar trick and extend 
$ D_m (| P'|, P) $ to complete integral over $ \mathbb{R}^d $ using the delta function
\begin{align*}
D_m(|P'|,P)= 2\int_{\mathbb{R}^d} dP^{''}\,  
&\frac{\delta({P^{''}}^2-P'^2)}{ |P^{''}|^{d-2} } \times\\
&\times t^c_{a_m}(k^{''}_{a_m},k_{a_m},k_{a_m}^2+i0)\delta(p^{''}_{a_m}-p_{a_m})G({\hat P^{''}})
H({P^{''}}^2).
\end{align*}
Here  $P'^2=k'^2_{a_m}+p'^2_{a_m}$, $P^{''}=\{k^{''}_{a_m} ,p^{''}_{a_m}\}$ and the auxiliary  
function of compact support $H$ is such that $H(P'^2)=1$. 
Calculation of the obtained integral using the delta functions included in the integrand leads to the following final result
\begin{align*}
&D_m(|P'|,P)= \frac{(P'^2-p^2_{a_m})^{(d_m-2)/2}}{|P'|^{(d-2)/2}} \times \\ 
&\times \int_{\mathbb{S}^{d_m-1}} 
d {\hat k^{''}_{a_m} } t^c_{a_m}\left((P'^2-p^2_{a_m})^{1/2} {\hat k^{''}}_{a_m},k_{a_m},k_{a_m}^2+i0\right)
G\left(\frac{(P'^2-p^2_{a_m}){\hat k^{''}_{a_m}}}{|P'|},\frac{p_{a_m}}{|P'|} \right), 
\end{align*}   
where  $d_m=3(N-m)$. 
So, we have shown that as a result of calculating the integral over the angular variables in 
(\ref {D-m}), the delta-functional singularities disappear. Applying this statement to the integral (\ref {Fpm}) taking into account the representations (\ref {TT} - \ref {tc}) for the 
$ T $-matrix with $ l = 1 $, we see that in the expressions for the quantities 
$ F^\pm (| P '|, P; G) $ from the first part of this section all delta-functional singularities disappear.

It remains for us to show that the pole singularities of the $ T $-matrix, described in Section 3, are smoothed by integration in the integral (\ref {Fpm}). In the general case, the constructions necessary to prove this fact turn out to be rather cumbersome and we restrict ourselves here to the case $ N = 3 $. As noted in section 3.2,
the $T$-matrix has two types of singular denominators, defined explicitly in (\ref {tatb}) and (\ref {FGJH}). In (\ref {denomin}) we showed that denominators of different types do not simultaneously vanish and therefore, when investigating the integral (\ref {Fpm}), they can be considered separately by choosing a suitable resolution of identity.
As a result of this procedure, singular integrals of two types will appear in the expression for the integral (\ref {Fpm})
\begin{align*}
I_1(|P'|,P) = \int_{\mathbb{S}^{d-1}}d{\hat P'} \, \frac{A_1(|P'|, {\hat P'},P)}
{k^2_b(p'_a,p_b)-k^2_b-i0} , \\
I_2(|P'|,P) = \int_{\mathbb{S}^{d-1}}d{\hat P'} \, \frac{A_2(|P'|, {\hat P'},P)}
{P^2+\lambda^2_a -{p'}^2_a+i0} ,  
\end{align*} 
where numerators of integrands collect contributions of smooth functions.  
Note that integrals of this type have arisen and have been studied in detail in Appendix III of \cite {Faddeev1963}. Here we give the key points in the proof of the smoothness of these integrals.

Let us consider the first integral  $I_1$. 
It is convenient to represent the measure of integration in the form
$$
 d{\hat P'}= (\cos\alpha)^2 (\sin\alpha)^2 \, d\alpha\,  d{\hat k'}_a\,  d{\hat p'}_a , 
$$
where $d{\hat k'}_a(d{\hat p'}_a)$  is the standard measure on the unit sphere  $\mathbb{S}^2$, and the hyperspherical angle $\alpha$ is defined in accordance with formulas 
\begin{equation*}
|k'_a|=|P'|\cos{\alpha}, \  \ |p'_a|=|P'|\sin\alpha.
\label{P'hs}
\end{equation*} 
Let us also use the standard representation for  $d{\hat p'}_a $ 
$$
d{\hat p'}_a =\sin \theta d\theta d\phi, 
$$
where the polar angle $\theta$  is measured from the vector  $p_b$, so that 
$$
\langle p'_a, p_b\rangle = |p'_a||p_b|\cos \theta. 
$$  
By the use of introduced variables, the integral $I_1$  takes the form 
\begin{align}
I_1(|P'|,P)= \int_{0}^{\pi/2} (\cos\alpha)^2 (\sin\alpha )^2
d\alpha \int_{\mathbb{S}^2} d{\hat k'}_a 
\int_{0}^{2\pi} d\phi \int_{0}^{\pi} \sin \theta d\theta \times \nonumber \\ 
\times \frac{A_1(|P'|,\alpha,{\hat k'}_a,\phi,\theta,P)}{\frac{p'^2_a}{s^2_{ab}} + \frac{c^2_{ab}p^2_b}{s^2_{ab}} - 2\frac{c_{ab}|p'_a||p_b|\cos\theta}{s^2_{ab}} -k^2_b -i0 }. 
\label{I1-hs}
\end{align}     
By substitution $ u = \cos \theta $, the integral over the variable $ \theta $
reduces to the standard singular integral
\begin{align}
\int_{0}^{\pi}\sin \theta d\theta ... = \int _{-1}^{1} du \frac{B_1(u)}{u-\zeta+i0} , 
\label{Int-u}
\end{align} 
where 
$$
\zeta = \frac{\frac{p'^2_a}{s^2_{ab}}+ \frac{c^2_{ab}p^2_b}{s^2_{ab}}    -k^2_b}{2\frac{c_{ab}|p'_a||p_b|}{s^2_{ab}}}, \ \ 
B_1=-\frac{A_1}{2\frac{c_{ab}|p'_a||p_b|}{s^2_{ab}}}. 
$$
Since $ B_1 (u) $ generally does not vanish at the ends of the integration interval $ (- 1,1) $, the integral on the right-hand side of (\ref {Int-u}) has logarithmic singularities, the explicit form of which is obtained by integrating in parts
\cite{Faddeev1963}
\begin{align}
\int _{-1}^{1} du \frac{B_1(u)}{u-\zeta+i0}= &\ln(1-\zeta-i0)B_1(1)-\ln(-1-\zeta-i0)B_1(-1) - \nonumber \\
&-\int_{-1}^{1}du \ln(u-\zeta-i0)\partial_u B_1(u). 
\label{I1-ln}
\end{align}
The integral on the right-hand side of (\ref {I1-ln}) converges absolutely and is a smooth function of $ \zeta $. Subsequent integration with respect to the variable $ \alpha $ in 
(\ref {I1-hs})
terms from (\ref {I1-ln}) with weak logarithmic singularities obviously leads to the fact that 
$ I_1 (| P '|, P) $ turns out to be a smooth function of its variables.

Consider further the integral $ I_2 $. In the same variables as $ I_1 $, this integral takes the form
\begin{align*}
I_2(|P'|,P)=\int_{\mathbb{S}^2} d{\hat k'}_a \int_{\mathbb{S}^2}d{\hat p'}_a \int_{0}^{\pi/2} d\alpha \, 
\frac{(\cos\alpha)^2(\sin\alpha)^2 \, A_2(|P'|,{\hat P'},P)}
{P^2+\lambda^2_a+i0 -|P'|^2(\sin\alpha)^2}. 
\end{align*}
Inner integral over the variable $ \alpha $ by substitution  $ v = (\sin  \alpha)^2 $
again reduces to the standard singular integral
\begin{align}
	\int_{0}^{\pi/2} d\alpha ... = \int_{0}^{1} dv \, \frac{\sqrt{v}\sqrt{1-v} B_2(v)}{v-z-i0}, 
\label{Int-alpha}
\end{align} 
with 
$$
z=\frac{P^2+\lambda^2_a}{|P'|^2}, \ \  B_2=-\frac{A_2}{2|P'|^2}. 
$$
The integrand in the integral on the right-hand side of (\ref {Int-alpha}) disappears at the ends of the interval of integration and, therefore, by virtue of Privalov's lemma \cite {Faddeev1963}, this integral is a smooth function of the variable $ z $ behaving like 
$ O ( z^{- 1}) $ for $ | z | \to \infty $. The latter prevents the appearance of the singularity at $ |P'| \to 0 $. As a result, we find that the original integral $ I_2 (|P'|, P) $ is a smooth function of its parameters.

So, we have shown that in the case $ N = 3 $, integration over the angular variables in the integral (\ref{Fpm}) leads to smoothing of all the singularities of the $T$-matrix. It can be shown that in the case $ N> 3 $ this statement will also be valid.


\section{Scattering problem in hyperspherical \\ representation}
 Hyperspherical representation for the Schr\"odinger equation 

\begin{equation}
({\bf H}_0 +{\bf V}-P^2) \Psi\equiv (-\Delta_X+V(X)-P^2)\Psi(X,P)=0
\label{SE}
\end{equation}
is obtained using hyperspherical coordinates for the vectors $ X $ of the configuration space
$\mathbb{R}_c^d$: $X=\{\varrho=|X|, {\hat X} \}$.  The Laplace operator in these coordinates is given by the expression  
\begin{equation*}
\Delta_X= \varrho^{-d+1}\partial_\varrho\varrho^{d-1}\partial_\varrho + \varrho^{-2}  
\Delta_{\mathbb{S}^{d-1}}, 
\label{H_0-hs} 
\end{equation*}
where $\Delta_{\mathbb{S}^{d-1}}$  is the Laplace-Beltrami operator on the unite sphere  $\mathbb{S}^{d-1}$. The operator ${\bf K }^2=- \Delta_{\mathbb{S}^{d-1}}$ 
in quantum mechanical applications is called the operator of squared hypermomentum. 
Its spectrum and the normalized eigenfunctions are well known satisfying the equation 
\begin{equation*}
{\bf K}^2 Y_{[K]} (\hat X)= K(K+d-2)Y_{[K]}(\hat X), 
\label{K2Y}
\end{equation*}       
where $K$ is non negative integer  and  $[K]$ denotes a multi-index, which includes, in addition to $ K $, a set of integer indices, the specific form of which depends on the choice of local coordinates on the unit sphere $\mathbb{S}^{d-1}$. The form of the eigenfunctions 
$ Y_ {[K]} (\hat X) $ also depends on this choice. A very general approach to the construction of these functions is given in \cite {Vilenkin1965} and the monograph \cite {Jib} contains many useful formulas for hyperspherical harmonics and literature references. In the hyperspherical representation, the wave function $ \Psi $ is given by an expansion in terms of the full set of hyperspherical functions
$Y_{[K]}(\hat X)\overline{Y}_{[N]} (\hat P)$ 
\begin{equation}
\Psi(X,P)= \sum_{[K]} \sum_{[N]}\Psi_{[K][N]}(\varrho,|P|) Y_{[K]}(\hat X) 
\overline{Y}_{[N]}(\hat P). 
\label{Psi-hs-exp}
\end{equation} 
The coefficients in the expansion (\ref {Psi-hs-exp}), called the partial components, are determined by the integrals
\begin{equation}
\Psi_{[K][N]}(\varrho,|P|) = 
\int_{\mathbb{S}^{d-1}} d{\hat P} \, Y_{[N]}(\hat P) \int_{\mathbb{S}^{d-1}} d{\hat X}\, 
\Psi(X,P)\overline{Y}_{[K]}(\hat X) . 
\label{PsiKN}
\end{equation}
By a standard technique, the Schr\"odinger equation (\ref {SE}) is reduced to the system of equations for the partial components
\begin{align}
\left\{ -\varrho^{-d+1}\partial_\varrho\varrho^{d-1}\partial_\varrho + \varrho^{-2} K(K+d-2) -P^2
\right\} \Psi_{[K][N]}(\varrho,|P|) =  \nonumber \\ 
-\sum_{[K']}
V_{[K][K']}(\varrho)
\Psi_{[K'][N]}(\varrho,|P|)  
\label{SE-hs}
\end{align} 
with potential matrix 
$$
V_{[K][K']}(\varrho) = 
\int_{\mathbb{S}^{d-2}} d {\hat X}\, \overline{Y}_{[K]}(\hat X) V(\varrho{\hat X}) Y_{[K']}(\hat X). 
$$

The scattering problem for the system of equations (\ref{SE-hs}) consists in finding a solution that satisfies certain asymptotic boundary conditions as $ \varrho \to \infty $ for the partial components of wave function (\ref{PsiKN}).
The most direct method for constructing these conditions is their derivation from the asymptotic properties
of the total wave function of a system of $ N $ particles (\ref {Psi}). The key means for this is the weak asymptotics (\ref{NBodyPsiWasymp}). 
Applying this formula to the inner integral over $\hat X $ with $ G = \overline {Y}_ {[K]} $ and then integrating over $ \hat P $ using the relation
$$
\int_{\mathbb{S}^{d-1}} d{\hat P}\, \overline{Y}_{[K]}(-\hat P)Y_{[N]}(\hat P) =
(-1)^K\delta_{[K][N]}, 
$$
we obtain the required result 
\begin{align}
\Psi_{[K][N]}(\varrho,|P|)    \sim    &N_d(|P|)  \times \nonumber \\ 
&\times \left\{   Q^-_d(|P|,\varrho) (-1)^K\delta_{[K][N]} -  
Q^+_d(|P|,\varrho) S_{[K][N]}(|P|)    \right\}. 
  \label{Psi-KN-as}    
\end{align}
Here the elements of the $S$-matrix $ S_ {[K] [N]} (| P |) $ according to (\ref {SS-N}) are given by the expression
\begin{align}
S_{[K][N]}(|P|)=\delta_{[K][N]}-
i\pi |P|^{d-2}\int_{\mathbb{S}^{d-1}}  d{\hat P}\, Y_{[N]}(\hat P) 
F^+(|P|,P;\overline{Y}_{[K]}). 
\label{S-KN}
\end{align}
The quantities $ S_{[K] [N]} (| P |) $ are formal matrix elements of the kernel (\ref {S-N}), however, since the $N$-particle  $T$-matrix contains singular terms, the calculation of the integrals in matrix elements requires care and should be done according to the formula (\ref {S-KN}).

Here it is appropriate to return to the question of asymptotic filtration. The leading terms of the asymptotics of the partial components in (\ref{Psi-KN-as}) contain elements of the $S$-matrix,
directly related to only one of the matrix elements  of the $N$-particle  scattering operator corresponding to transitions between the scattering channels, in which all particles move asymptotically freely. The kernel of this element of the scattering operator
is expressed in terms of the $T$-matrix by the formula \cite{MerkFadd, Faddeev1963}
\begin{equation}
S_{00}(P',P)= \delta(P'-P)-2\pi i \delta(P'^2-P^2)T(P',P,P^2+i0). 
\label{S00}
\end{equation}  
An integral operator with kernel (\ref{S00}) acts on an arbitrary function by the formula
\begin{align*}
\int_{\mathbb{R}^d}dP'\, & S_{00}(P',P) {\hat f}(P')=
\nonumber \\ 
&\int_{\mathbb{S}^{d-1}} d{\hat P'}\, 
[\delta({\hat P'},{\hat P}) -i\pi  |P|^{d-2} T(|P|{\hat P'},P,P^2+i0)  ] {\hat f}(|P|{\hat P'}). 
\end{align*}
The kernel in square brackets is called the restriction of $S_ {00}$ to the energy surface and coincides with the kernel (\ref {S-N}), whose matrix elements form the matrix  
$S_ {[K] [N]}$.

We have completed the formulation of the scattering problem for a system of $N$ particles in a continuum in  hyperspherical representation by formulating equations for the partial components of the wave function (\ref {SE-hs}) and asymptotic boundary conditions (\ref {Psi-KN-as}).

\section{Conclusion}
We have obtained weak asymptotics of the wave functions of quantum systems of $N$ particles corresponding to the initial states with free particles. In contrast to the coordinate asymptotics of such wave functions in the complete configuration space, which have an extremely complex structure, the leading terms of weak asymptotics are described by a superposition of standard incoming  and outgoing  spherical waves of the corresponding dimension. The coefficients of this superposition are the amplitudes expressed in terms of one of the matrix elements of the total 
$N$-particle scattering operator corresponding to transitions between asymptotic states in which all $N$ particles are free both before and after the interaction. The processes of formation of bound states of subsystems (reactions) and processes of rescattering of particles do not contribute to the leading terms of weak asymptotics. This phenomenon of asymptotic filtering, previously unknown, allows us to formulate correct asymptotic boundary conditions for the partial components of the $N$ particle wave function in the hyperspherical representation widely used in applications. The latter makes it possible to abandon the phenomenological concept of "democratic" scattering \cite {Jib}, within the framework of which it was assumed that in the process of collisions only elastic processes are possible without reactions and consecutive
rescattering of subsystems. Although these assumptions obviously cannot be physically substantiated, they had to be used for quite a long time in applications due to the lack of exact and correct asymptotic representations for the partial components of wave functions obtained from first principles. The results of this work eliminate this circumstance and will be useful, in particular, for the practical solution of the scattering problem for systems of $N$ particles in the hyperspherical representation.

\section*{Acknowledgement
}
The work has been  supported by the RFBR grant  18-02-00492 а. 




\centerline{---------------} 

\end{document}